\documentstyle[preprint,aps]{revtex}

\begin{document}
\title{\bf Universality Classes in Isotropic,  Abelian and non-Abelian, Sandpile Models}
\author{Erel Milshtein\footnote{E-mail: milstein@flounder.fiz.huji.ac.il},
Ofer Biham\footnote{E-mail: biham@cc.huji.ac.il}
and Sorin Solomon\footnote{E-mail: sorin@cc.huji.ac.il}}
\address{
Racah Institute of Physics, 
The Hebrew University, 
Jerusalem 91904, 
Israel}

\maketitle

\begin{abstract}
Universality in isotropic,  abelian and non-abelian, sandpile models is examined using extensive
numerical simulations. To characterize the critical behavior we employ an extended set of 
critical exponents,  geometric features of the avalanches,  as well as scaling 
functions describing the time evolution of average quantities such as the area and size
during the avalanche.
Comparing between the abelian Bak-Tang-Wiesenfeld model [P. Bak,  C. Tang and K. Wiensenfeld, 
 Phys. Rev. Lett. 59,  381 (1987)],  and the non-abelian models introduced by Manna [S. S. Manna,  J. Phys. A. 24,  L363 (1991)]
and Zhang [Y. C. Zhang,  Phys. Rev. Lett. 63,  470 (1989)] we find strong indications that each one of
these models belongs to a distinct universality class.
\end{abstract}


\newpage

\section{Introduction}
	\label{intro.}

Sandpile models have been studied extensively during the last decade as a paradigm of self-organized
 criticality (SOC). This concept,  introduced by Bak,  Tang and Wiesenfeld (BTW)~\cite{B87,B88}, stimulated 
numerous theoretical~\cite{T88,D89,H89,H89a,D90,C90,G90,C90a,G91,P91,L91a,P96},  
numerical~\cite{Z89,K89,G90a,M90,M90a,M91a,C91,C93}
and experimental studies~\cite{R90,H90,L91,N92,C94,F96}.
In sandpile models,  which are defined on a lattice,  grains are 
deposited randomly until the height at some site exceeds a 
threshold,  and becomes unstable.  
Grains are then distributed to the nearest neighbors.  
As a result of this relaxation process neighboring sites may become
 unstable,  resulting in a cascade of relaxations called {\it an 
avalanche}.
It was observed that these models are self-driven into a critical 
state which is characterized by a set of exponents~\cite{B87,B88}.
This set includes {\it distribution exponents} that describe the distribution of 
quantities such as avalanche size,  area,  and lifetime,  and the {\it geometric exponents} which 
relate various properties of the dynamics~\cite{C91}.
Additional sandpile models which differ from the BTW model in the dynamic rules 
were introduced and studied.
These include the Manna model~\cite{M90,M91a},  in which the dynamics during an 
avalanche is
stochastic and the Zhang model~\cite{Z89},
in which the dynamic variable is continuous.
In order to understand the basic mechanism of SOC,  numerous 
attempts were made to assign the various models to universality classes. 
To this end,  Vespignani, Zapperi and Pietronero introduced the fixed scale transformation
which is a real space renormalization group (RG) approach~\cite{V94,V95}.
They applied this approach to the BTW and Manna models and concluded that these two models
belong to the same universality class.
Later, a comparison between the critical behavior of
BTW and the Manna models was performed, using extensive numerical simulations
and an extended set of exponents~\cite{B96}. The results provide strong evidence that the 
two models belong to different universality classes. 

Recently, Diaz-Guilera and Corral applied the dynamic RG technique to study critical behavior in sandpile 
models. Their conclusion was that the BTW and the Zhang 
models belong to the same universality class~\cite{D92,D94,C97}. 
A similar conclusion was reached by L\"{u}beck~\cite{L97}, based on numerical simulations.

In this paper we apply an extended set of tools for the characterization of critical behavior in sandpile
models and for their classification into universality classes. These tools include measures that characterize 
avalanches as a whole such as the distribution exponents, the geometric exponents and the avalanche structure.
We also introduce a new measure,
based on scaling functions, describing the time evolution of the area, size and energy during an avalanche.
Combining all these tools, we find strong indications that the BTW, Manna and 
Zhang models belong to three different universality classes.

The paper is organized as follows. The models are introduced in Sec. II and the measures for
their classification are presented in Sec. III. The simulations and results are given in Sec. IV, followed by
 a summary and conclusions in Sec. V. 

\section{The Models}
	\label{models}

Sandpile models are defined on a $d$ dimensional lattice of linear size 
$L$.
Each site {\bf i} is assigned a dynamic variable $E(\bf i)$ which 
represents some physical quantity such as energy,  stress etc.
In a critical height model a configuration
$\left \{E({\bf i}) \right \}$ is called {\it stable} if
for all sites $E({\bf i}) < E_c$,  where $E_c$ is a threshold value.
The evolution between stable configurations is by the following 
rules:

 (i) Adding energy.
Given a stable configuration $\{E(\bf j)\}$ we select a 
site ${\bf i}$ at random and increase $E({\bf i})$ by some amount 
$\delta E$.
When an unstable configuration is reached rule (ii) is applied.

(ii) Relaxation rule.
If $E(\bf i) \geq$ 
$E_c$,   
relaxation takes place and energy is distributed in the following way:

\begin{eqnarray}
\label{def}
E({\bf i}) & \rightarrow & E({\bf i}) - \sum_{\bf e}\Delta 
E({\bf e})  \nonumber \\
\\				
E({\bf i}+{\bf e}) & \rightarrow & E({\bf i}+{\bf e})+
\Delta E({\bf e}), \nonumber 
\end{eqnarray}
where ${\bf e}$ are a set of (unit) vectors from the site ${\bf i}$
to some neighbors. 
As a result of the relaxation the dynamic variable in one or more of
 the neighbors may exceed the threshold.  
The relaxation rule is then applied until a stable 
configuration is reached.
The sequence of relaxations is an avalanche which propagates 
through the lattice.

It was shown before, that the parameters $\delta E$ and $E_c$  are irrelevant to the scaling 
behavior~\cite{D90,D92,D94}.
This indicates that the critical exponents depend only on the
vector $\Delta E$,  to be termed {\it relaxation vector}. 
For a square lattice with relaxation to nearest neighbors (NN) it is of 
the form $\Delta E=(E_N, E_E, E_S, E_W)$,  where $E_N, E_E,  E_S$ and $E_W$ are
the amounts transferred to the northern,  eastern,  southern and western NN's respectively. 
In the BTW model, $E_c=4$, $\delta E=1$ and $\Delta E=(1, 1, 1, 1)$. If an active 
site with  $E({\bf i}) > E_c$ is toppled, it would not become empty after 
the topple had occurred. 
In the Zhang model~\cite{Z89}, for which $E_c=1$ and $0<\delta E<1$, the relaxation vector is given by $(b, b, b, b)$, 
where $b=E({\bf i})/4$ and $E({\bf i})$ 
is the amount of energy in the active site before
the topple had occurred . Obviously,  the site ${\bf i}$
remains empty after toppling.

In a random relaxation model~\cite{M91} a set of neighbors is randomly chosen 
for relaxation.
Such a model is specified by a set of relaxation vectors,  each 
vector being assigned a probability for it application.
For example,  a possible realization of a two-state model includes six relaxation 
vectors (1, 1, 0, 0), (1, 0, 1, 0), (1, 0, 0, 1), (0, 1, 1, 0), (0, 1, 0, 1) and (0, 0, 1, 1),  each one applied with a 
probability of $1/6$. A four state model  would include relaxation vectors such as (4, 0, 0, 0), (3, 1, 0, 0), 
(2, 2, 0, 0), (2, 1, 1, 0), (1, 1, 1, 1), (0, 2, 1, 1), etc., applied at
different probabilities while maintaining the four fold symmetry of the relaxation rule.
A {\it time step} (of unit {\it time}) is defined as the relaxation of all the
sites having  $E({\bf i}) \geq  E_c$,  
after the completion of the previous time step. 
A model is said to be {\it abelian} if the configuration after the avalanche,  does not depend 
on the order in which the relaxation of the active sites was performed.
The BTW model was shown to be abelian~\cite{D90}.
The Manna models~\cite{M90,M91a} are not abelian because they contain a 
random choice of the toppling direction. As a result,  they develop different scenes of toppling 
that depend on the order of relaxation of the active sites in a single time step. The Zhang
model is also non-abelian. This can be seen from the following example,  when two active NN 
 sites are toppled within the same time step. The site that was toppled last remains empty while 
the other one is non-empty. This shows that the final configuration depends on the order.

\section{Measures for classification}
	\label{MFC}

Avalanches have various properties which can be measured in a 
simulation:
size,  area,  lifetime,  linear size,  and perimeter.
The size ($s$) of an avalanche is the total number of relaxation 
events that occurred in the course of a single avalanche.
The area ($a$) is the number of sites in the lattice where 
relaxation occurred.
The lifetime ($t$) of an avalanche is the number of time steps (defined above) that took place during
the avalanche.
As for the linear size of an avalanche,  
one can use the radius of gyration ($r$) of the cluster 
of sites where relaxation occurred.
A site inside the area $a$ is
 defined as a perimeter site if it has a nearest neighbor where 
no relaxation took place.
The perimeter ($p$) is the number of perimeter sites. 
Thus we have a set of variables $\{s, a, t, r, p\}$ which 
characterize an avalanche.
The avalanche variables have probability functions which are assumed
 to fall off with a power law defined by
$P(x) \sim x^{1-\tau_{x}}$, 
where $x \in \{s, a, t, r, p\}$ and the exponents $\tau_{x}$ are called {\it distribution exponents}.
These variables also scale against each other in the form
\begin{equation}
	y \sim x^{\gamma_{yx}}, 
\end{equation}
for $x, y \in \{s, a, t, r, p\}$ and the exponents $\gamma_{yx}$ are 
called {\it geometric exponents}.
The exact definition of the geometric critical exponents $\gamma_{xy}$ is in terms of conditional 
expectations values: 
$E[y | x] \sim x^{\gamma_{yx}}$~\cite{C91}.
The exponents are not independent.
It is shown in~\cite{C93} that they satisfy the scaling relations
\begin{equation}
\label{gamma}
\gamma_{yx}=\gamma_{xy}^{-1}
\end{equation}
and
\begin{equation}
\gamma_{zx}=\gamma_{zy}\gamma_{yx}.
\end{equation}
Apart from looking at the critical exponents we have also examined the structural
features of the avalanches. We define on the lattice a function that specifies the 
number of toppling events,  $f(\bf i)$,  
 for each site ${\bf i}$ 
  during a single avalanche.
 In a two dimensional (2D) model this function takes the form of a hilly terrain, with discrete 
heights.
 The $n$th terrace is 
 then defined as the set of sites for which $f({\bf i})$
 $ \geq n$.
 This height profile of an avalanche can be described by drawing the area of the 
   $n$th terrace, $A_n$ vs. $n$. Furthermore,  we can 
consider $A_n$ as an
 avalanche variable,  just like the size or the area. This gives us  
 new variables that can be measured and compared
 between the various models . 

 The avalanche properties introduced so far, such as the area and size, characterize an 
 avalanche as a whole, and are measured only after the avalanche is completed.
 Here we introduce measures to characterize the time evolution during an avalanche.
 Measured vs. time, for a single avalanche, these measures exhibit an irregular form.
 However, if we average them over a large number of avalanches, a typical shape emerges.
 We now introduce three such measures:

\begin{enumerate}
\item The avalanche area $a_c(t)$, 
namely the number of sites where at least one relaxation occurred
during the first $t$ time steps of the avalanche. As the avalanche
   is completed  $a_c(t)$ evolves to the area $a$ of the avalanche.
We also define $a(t)$ as the time derivative of $a_c(t)$, according to $a(t)=da_c(t)/dt$
where $da_c(t)/dt \equiv a_c(t+1)-a_c(t)$.
The variable $a(t)$ gives the number of sites that at time $t$ became active for the 
first time (and are to be toppled in the next time step).
As the avalanche evolves to an end we find that the avalanche area satisfies 
$a=\sum_{t=0}^{t_{max}} a(t)$, where  $t_{max}$ is the avalanche time, and the $t=0$
step consists of the deposition event that initiated the avalanche.
\item The number of active sites $s(t)$ (namely sites that are to topple in the next time
   step),  as a function of time. As the avalanche evolves to an end we find
   that $s=\sum_{t=0}^{t_{max}} s(t)$ is the avalanche size. 
\item Considering a sandpile in a gravitational field,  we define the potential energy 
$U(t) = \sum_{\bf i} u_{\bf i}(t)$, where $u_{\bf i}(t)$ is the energy at site ${\bf i}$ given by :
\begin{equation}
u_{\bf i}(t) = \int_{0}^{E(\bf i)} mgh(t) dh .
\end{equation} 
   The values of $m$
   and $g$ are irrelevant and we take $mg=1$.
   The potential energy defined here,  turns out to be proportional to $E(\bf i)^2$, 
   contrary to the ordinary definition of energy in sandpiles, which is linear in $E({\bf i})$. 
   For the case of a discrete dynamic variable, a site ${\bf i}$ having energy $E({\bf i})$ will
   have a potential energy 
   $u_{\bf i}(t) = E({\bf i})(E({\bf i})+1)/2$. 
\end{enumerate}

Averaging the functions $a(t)$, $s(t)$ and $u(t)$ over a large number of avalanches we
obtain the functions $A(t)$, $S(t)$ and $U(t)$, respectively.
According to the dynamic scaling assumption, each one of these functions can be
written in the general scaling form :
\begin{equation}
X(t) = K_X\langle t \rangle_X^{-\alpha_X} f_X\left(\frac{t}{\langle t \rangle_X} \right )
\end{equation} 
where 
\begin{equation}
\langle t \rangle_X=\frac{\sum_{t} tX(t)}{\sum_{t} X(t)}
\end{equation} 
and $X \in \{U, S, A\}$.
The scaling function $f_X(\mu)$, where $ \mu=t/ \langle t \rangle_X$ satisfies the sum rules 
\begin{equation}
\int_0^{\infty} f_X(\mu) d\mu = \int_0^{\infty} \mu f_X(\mu) d\mu = 1.
\end{equation} 
The shape of the scaling function, and the values of the exponents $\alpha_X$
can be used to distinguish between universality classes.
Moreover, the relation between these scaling functions can be used as a further tool. 
For example, if  $f_X(\mu)$ and $f_Y(\mu)$ coincide for one model and are different 
in another model, it indicates that these two models do not belong to the 
same universality class.
The dependence of $\langle t \rangle_x$ on the system size $L$ is given by 
\begin{equation}
\langle t \rangle_X \sim L^{\beta_X}.
\end{equation} 

\section{Simulations and Results}
	\label{simul.}

Having defined the three models and the measures used for their characterization, 
we now describe the computer simulations.
We have used open boundary conditions and system sizes up to $512^2$,  
with $10^6$ to $10^8$ grains dropped,  in two dimensions (2D). 
For each run we ascertained, before collecting data, that the dynamics has reached the
 critical state by applying Dhar's burning algorithm~\cite{D90},  or by starting
 with a configuration belonging to the critical state.
For each model we have calculated all the measures of classification mentioned in Sec. III.

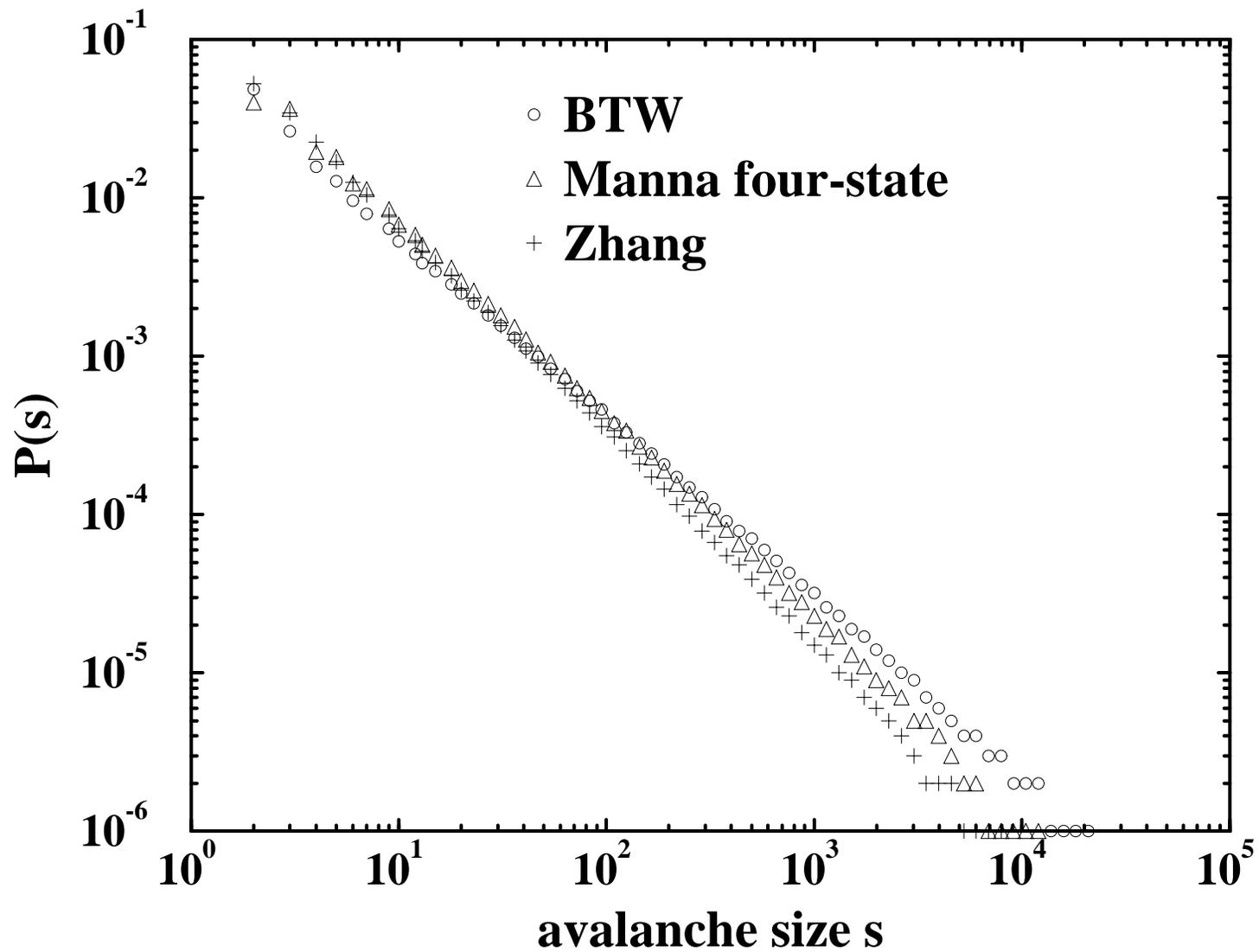
\begin{figure}
\caption{Avalanche size distributions in the BTW,  Manna four-state and 
Zhang models. The exponents are $\tau_s = 2.09 \pm 0.005  $ for BTW,  $\tau_s = 2.23  \pm 0.01  $ for 
Manna and  $\tau_s = 2.25  \pm 0.01 $ for the Zhang model.
 System size is $128^2$.}
\label{taus}
\end{figure}

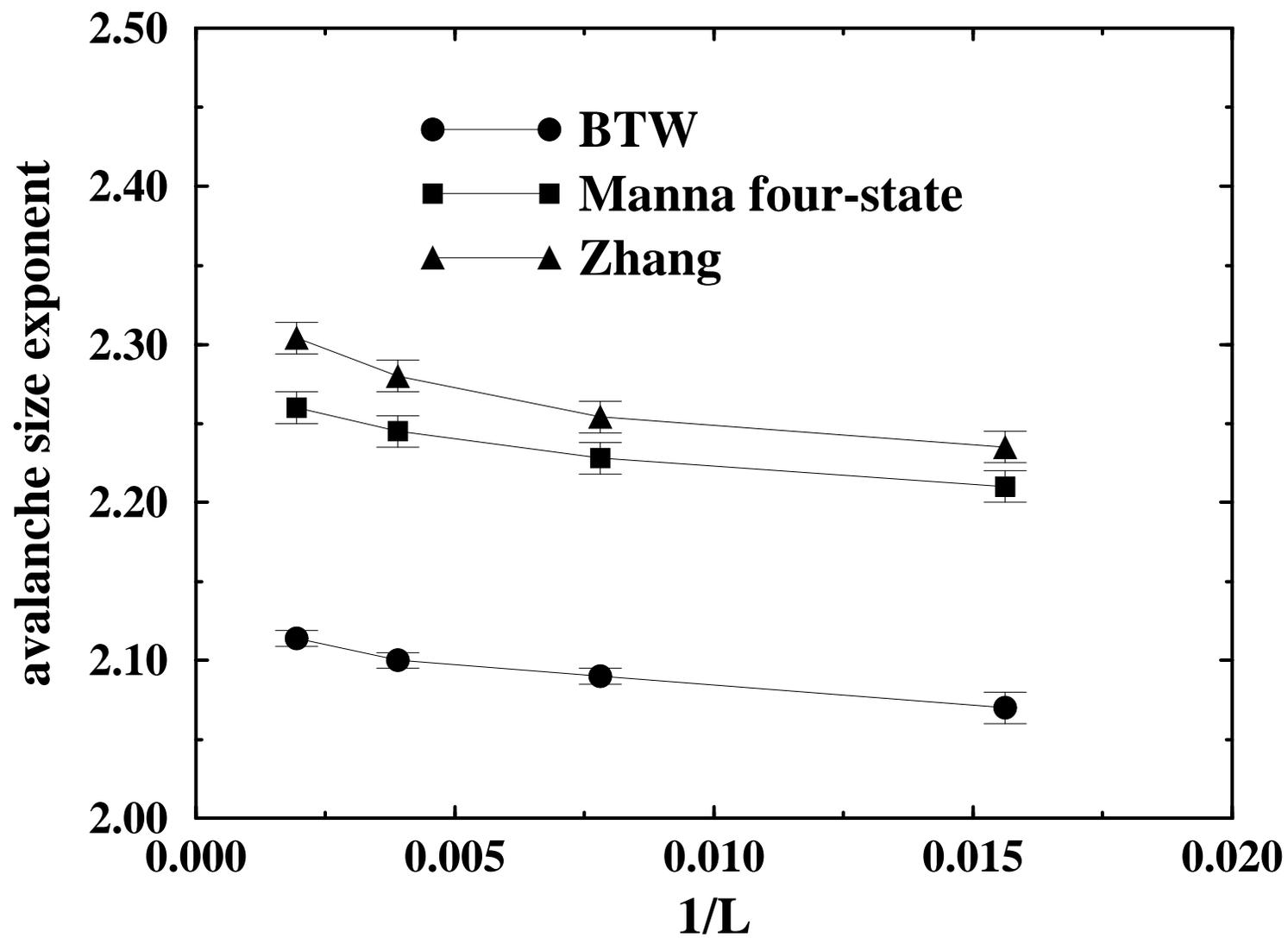
\begin{figure}
\caption{The exponent $\tau_s$ of the avalanche size distribution
as a function of the inverse system size ($1/L$) in the BTW,  Manna four-state and 
Zhang models. The lines are guides to the eye. Although it is hard to reliably extrapolate from
these results to $1/L \rightarrow 0$, this graph strongly indicates that the curves converge to 
three different values.}
\label{tau_size}
\end{figure}

The exponent $\tau_s$ that describes the avalanche size distribution was measured for  the BTW,  
Manna four-state and Zhang models  (Fig.~\ref{taus}) finding good agreement with previous 
results ~\cite{B96,L97}. 
The exponent $\tau_s$ for the three models, as a function of the inverse system size, is given in 
 Fig.~\ref{tau_size}. These results do not provide a reliable extrapolation of $\tau_s$ to the 
infinite system limit.
However, they strongly indicate that the curves converge to different values of $\tau_s$ as $L \rightarrow \infty$.
As $\tau_s$ exhibits relatively strong dependence on the system size~\cite{B96}, it cannot be used as 
the primary tool for classification of models, but only to provide additional evidence.

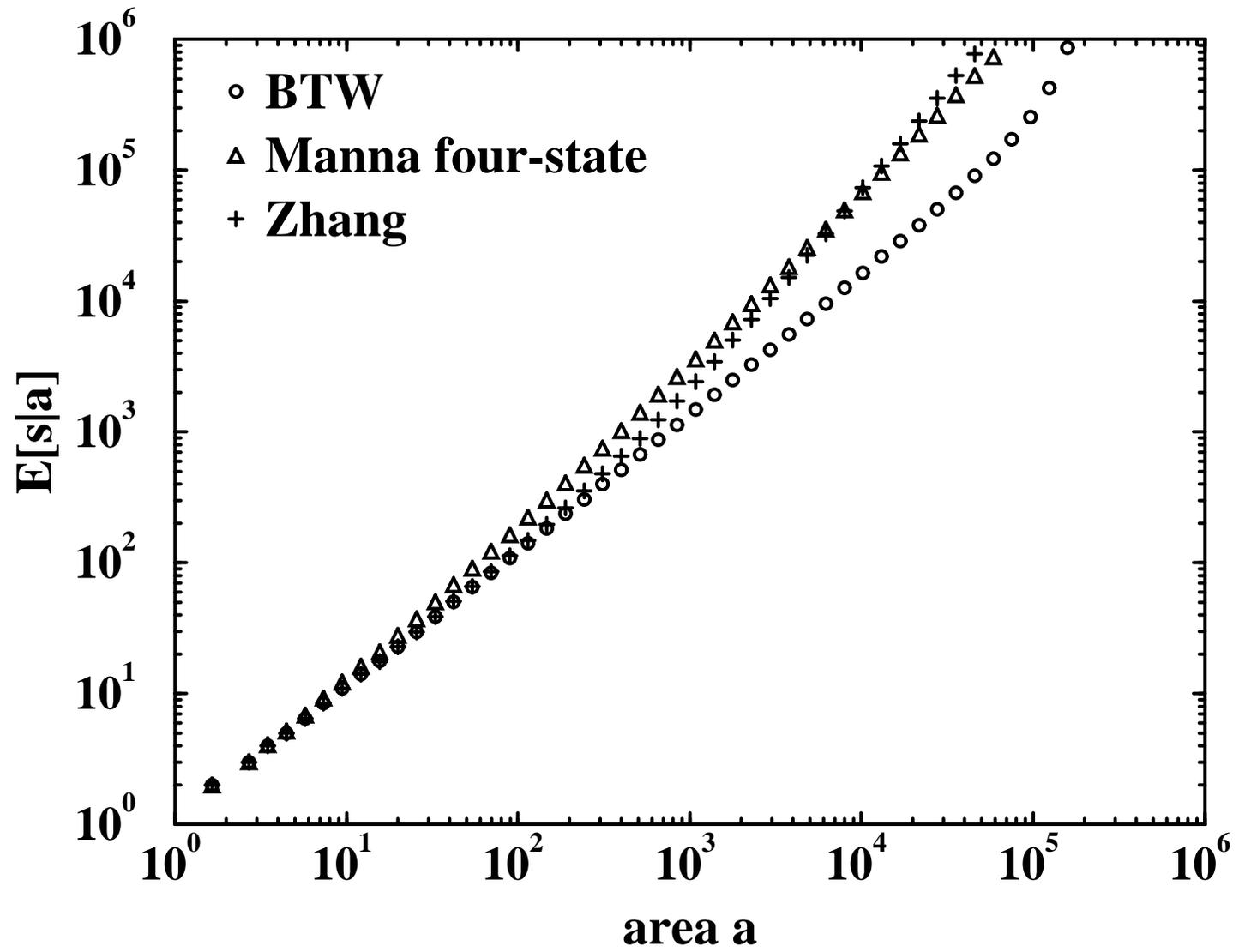
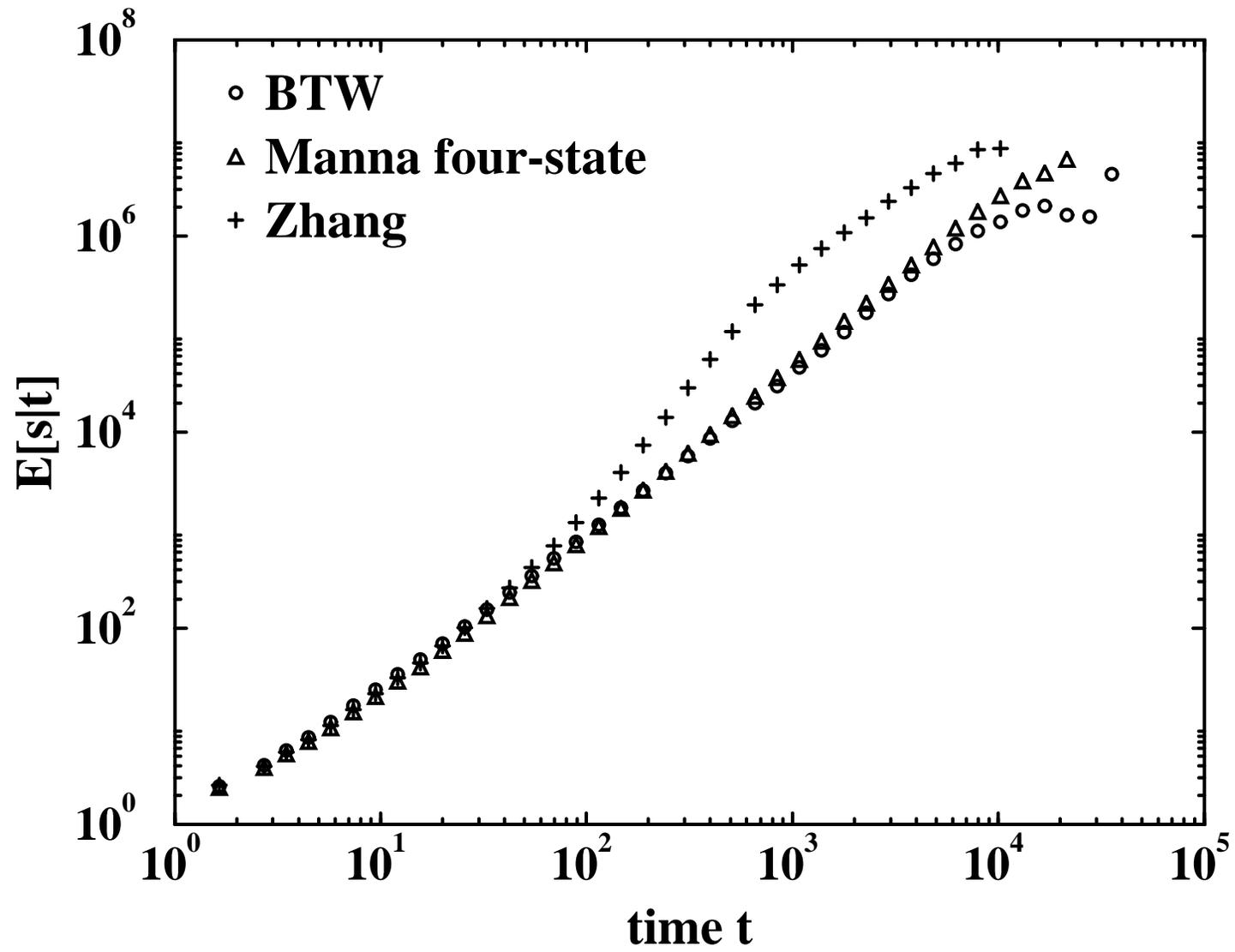
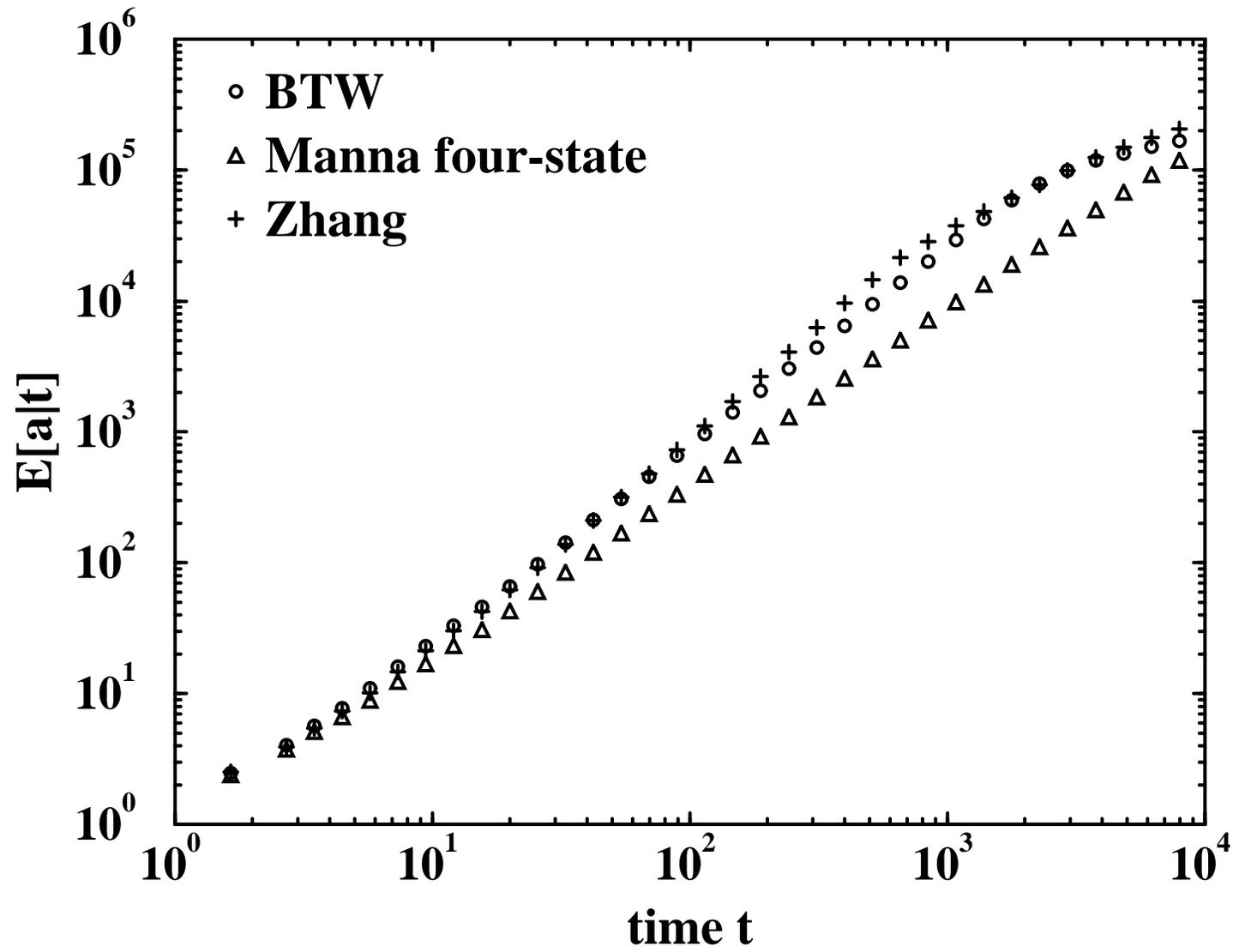
\begin{figure}
\caption{Geometric critical exponents for the BTW, Manna four-state and Zhang models.
(a) $E[s|a]$ (average avalanche size for given avalanche area) vs. $a$, is presented, yielding $\gamma_{sa}$.
(b) $E[s|t]$ is given  vs. $t$ (avalanche time), yielding $\gamma_{st}$.
(c) $E[a|t]$ is given  vs. $t$, yielding $\gamma_{at}$.
System size is $512^2$ with $10^7$ grains dropped.
Data were binned with bin size increasing exponentially.}
\label{gammah}
\end{figure}

The geometric exponents  $\gamma_{xy}$ are only weakly dependent on the system size,
and turn out to be very useful for classification of sandpile models~\cite{B96}.
In our simulations we examined the  exponents $\gamma_{sa},  \gamma_{st}$ and $\gamma_{at}$.
This was done by drawing on a log-log scale quantities such as the average avalanche size 
$E[s|a]$ for a given area $a$, where  $\gamma_{sa}$ is given by the slope of the 
straight line section (Fig.~\ref{gammah}).
 The values of the geometric
exponents for the BTW  and the Manna four-state models,  are in agreement with previous simulations~\cite{B96}. 
Our simulations of the Zhang model, for system size up to $512^2$,
showed that these variables,   $\gamma_{sa},  \gamma_{st}$ and $\gamma_{at}$, 
{ \em are not scale invariant},  (Fig.~\ref{gammah}). Although for small avalanches 
the scaling behavior for the Zhang model 
 resembles the distribution of the same variables in the BTW 
model,  for large avalanches,  we find a {\it bend} 
and a sudden change in slope.
This bend is seen for all the system sizes that were checked.
 For large avalanches, the geometric
exponents are clearly different from the values recorded for the BTW model and  for the 
Manna four-state models.  This puts in question the previous
assignment of the Zhang model to the universality class of the 
BTW model~\cite{D92,D94,L97,P93}.

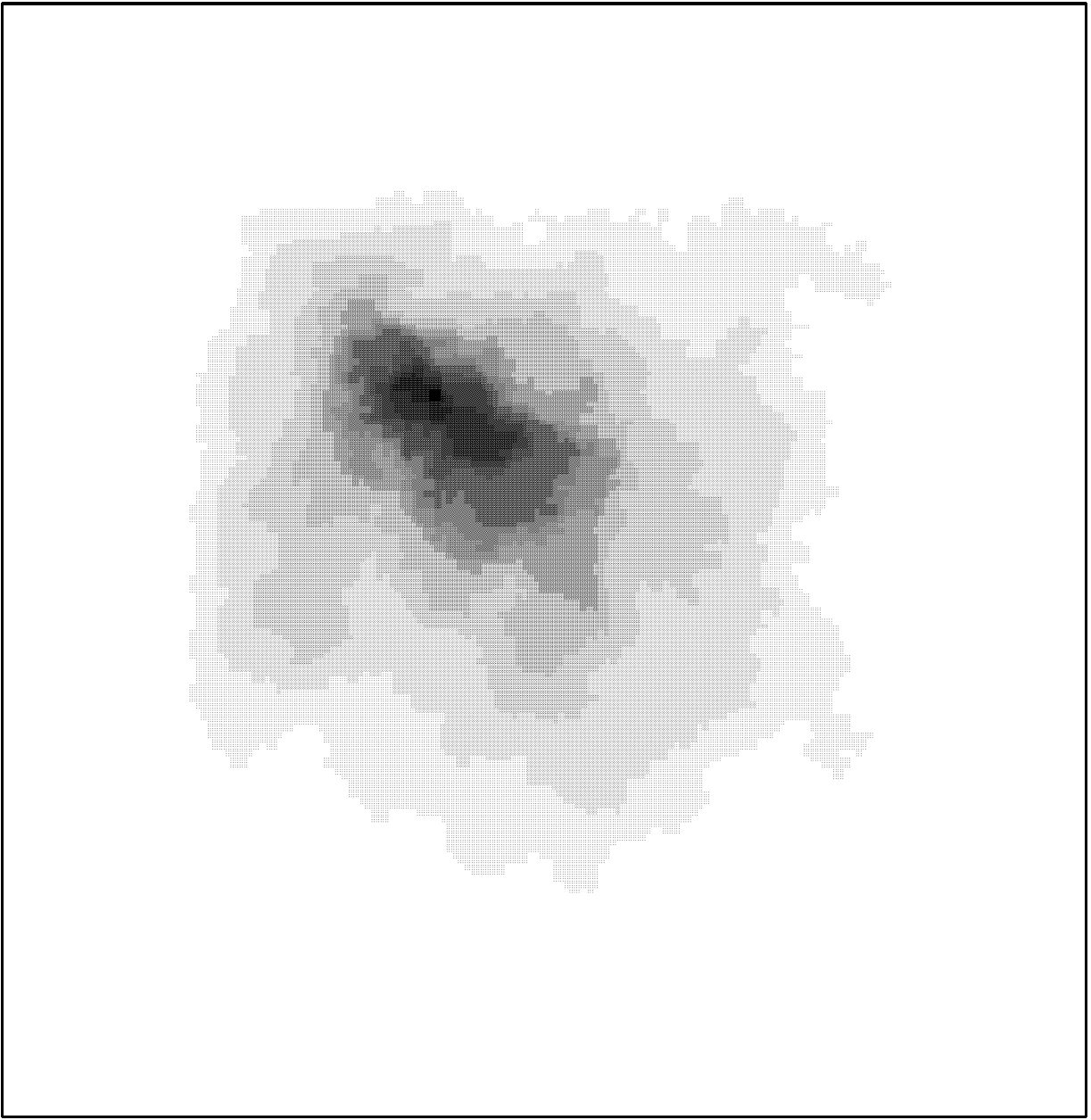
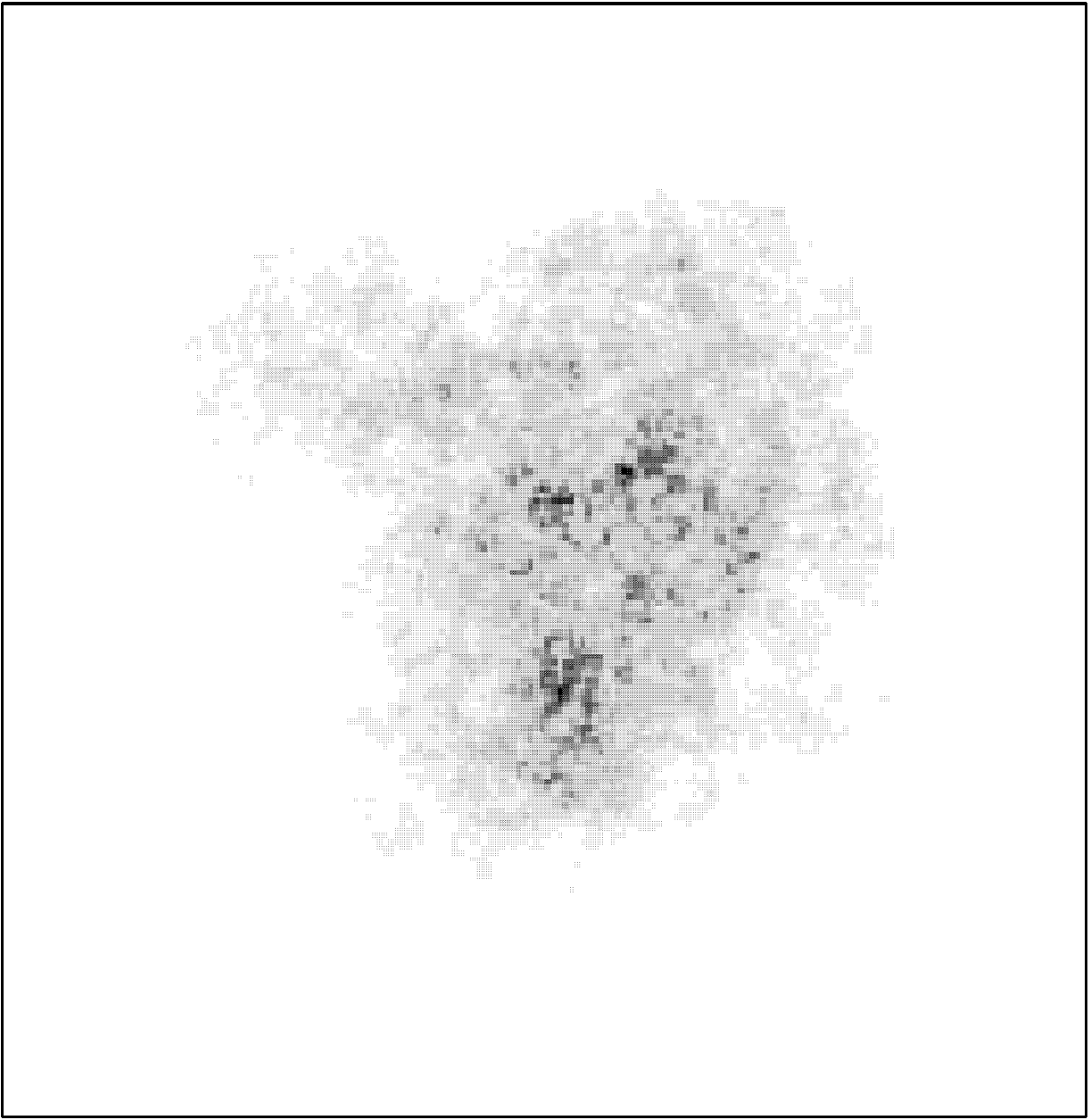
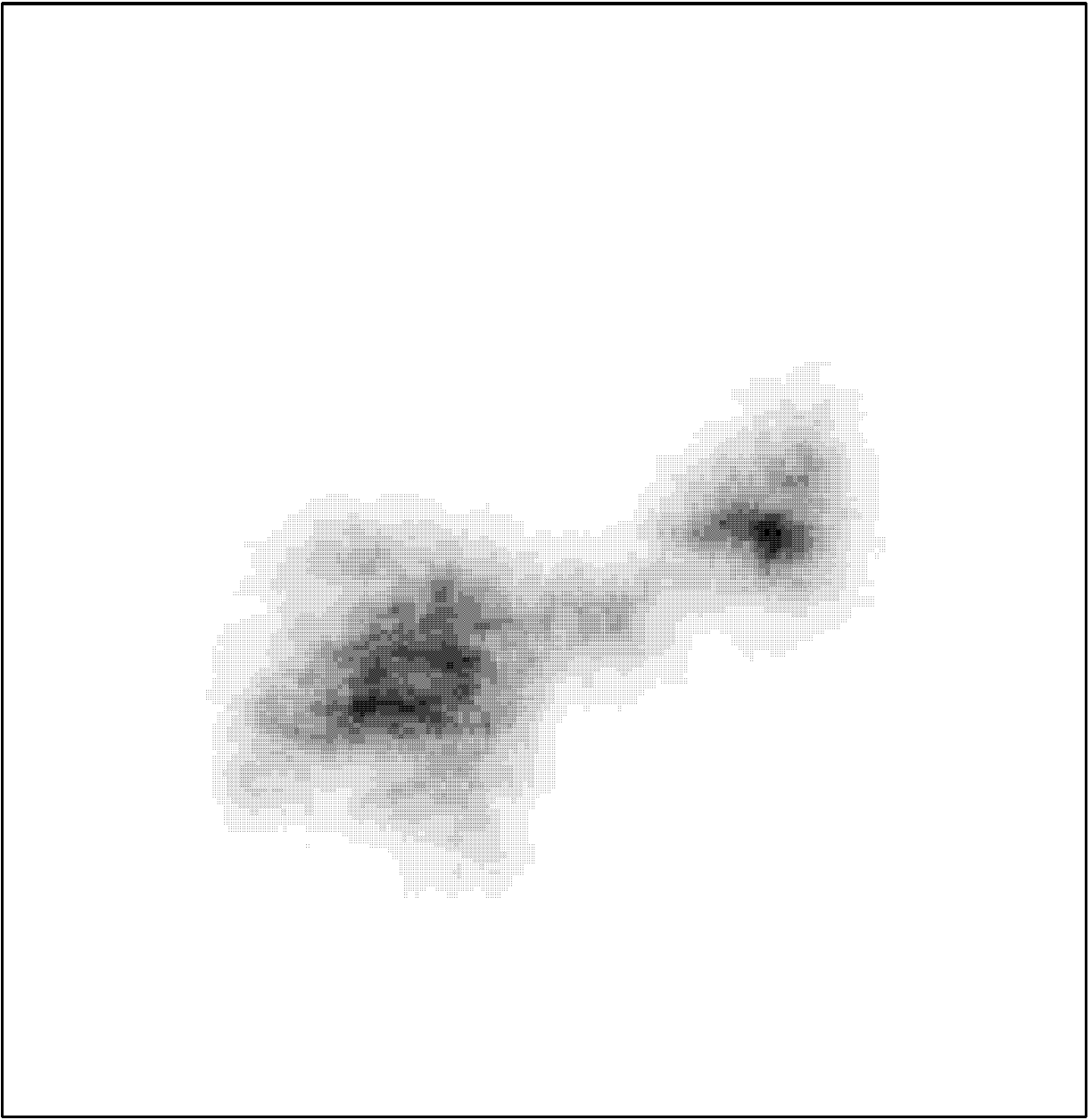
\begin{figure}
\caption{Typical large avalanche structure for the BTW model (a),  
 Manna four-state model (b) and the  Zhang model (c). Gray scales indicate the number of 
toppling events $f({\bf i})$ which occurred at each site during the avalanche.
White represents zero relaxations,  and black represents the 
maximal no. of relaxations [17 in (a),  34 in (b) and 22 in (c)].
System size is $128^2$.
Note the shell structure in the BTW avalanche~\protect\cite{D89} vs. the irregular structure of the avalanche 
in the Manna four-state model, and the intermediate  structure of the Zhang avalanche.
These qualitative geometrical differences translate into 
quantitative differences in exponent values.}
\label{largeavalanche}
\end{figure}

To further characterize the avalanche structure we examined the function $f({\bf i})$,
that provides the number of toppling events at site ${\bf i}$ during the avalanche (Fig.~\ref{largeavalanche}).
For the BTW model,  we observe a shell structure in which all sites which relaxed at least $n+1$ 
times form a connected cluster with no holes which is contained in
 the cluster of sites which relaxed at least $n$ times~\cite{D90,I94}.
The Manna four-state model exhibits a random avalanche structure 
with many peaks and holes~\cite{B96}. In between we find the Zhang model,  
which shows an avalanche structure which is mostly shelled,  but is different from the BTW picture by 
having several peaks and holes,  but not as many as in the Manna four-state model. 

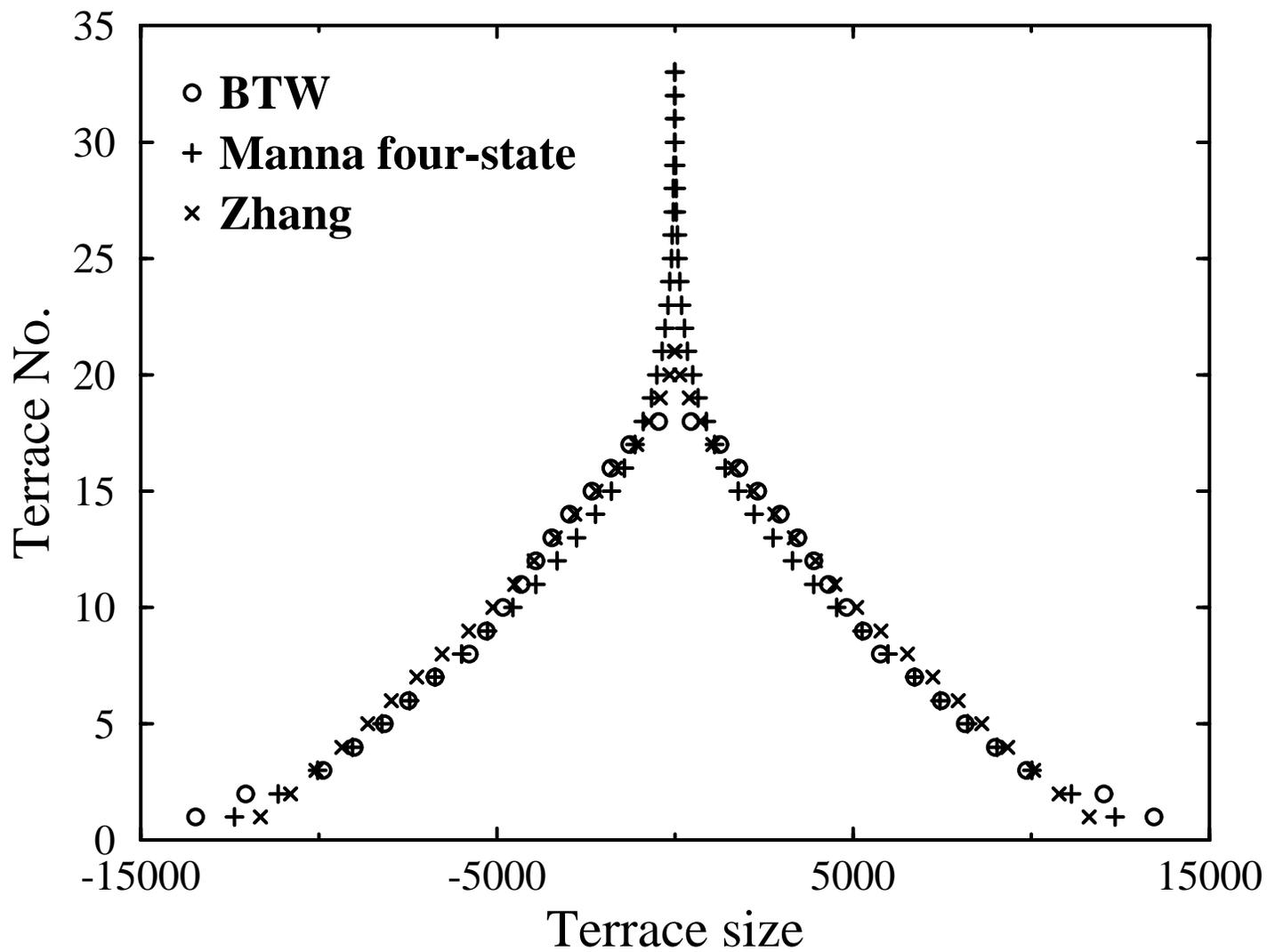
\begin{figure}
\caption{The activity profile of a typical large avalanche in the BTW, Manna four-state and Zhang
model. The terrace number is plotted as a function of its area. The picture is reflected around the $y$
axis. The system size is $128^2$.
The BTW and Zhang models exhibit moderate slopes, while in the Manna four-state model the slope
becomes extremely steep at high terrace numbers.}
\label{terraces}
\end{figure}

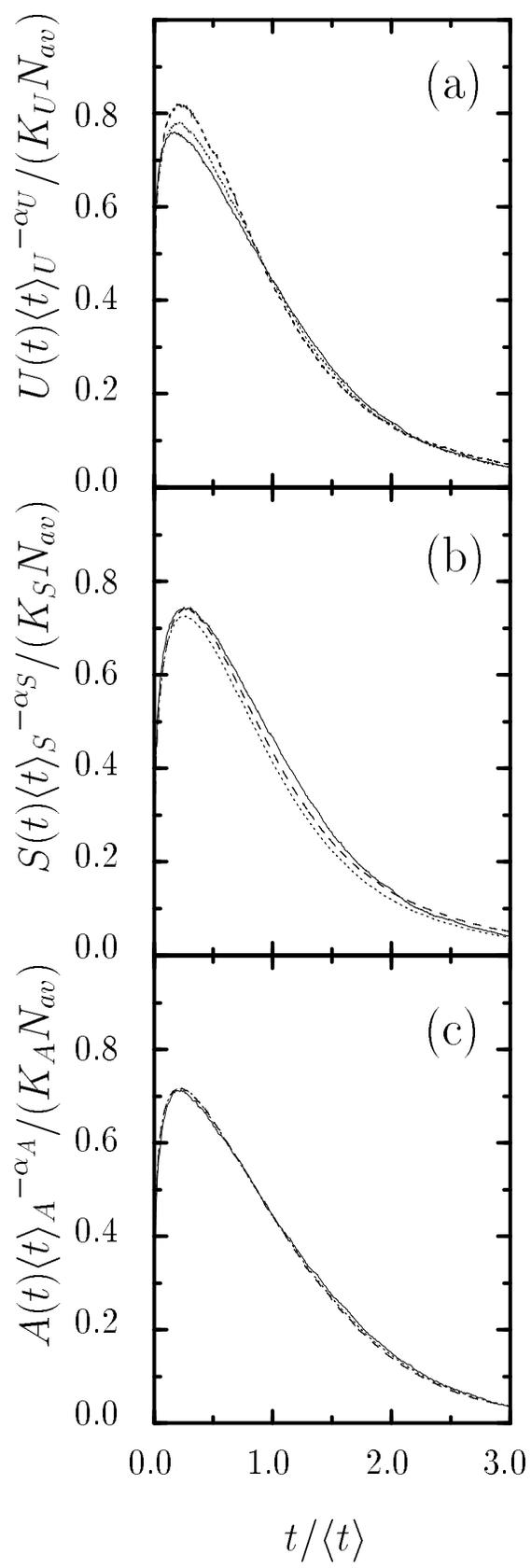
\begin{figure}
\caption{The scaling functions for the BTW model for $system size=128,256$ and $512$
(a) $f_U(\mu)$ which describes the time dependence of 
the energy during the avalanche,
(b) $f_S(\mu)$ which describes the time dependence of the number of active sites
and (c) $f_A(\mu)$ which describes the time dependence of the avalanche area growth.
 We observe that all three scaling functions
coincide, indicating a common scaling function for $U,S$ and $A$.}
\label{btwscale}
\end{figure}

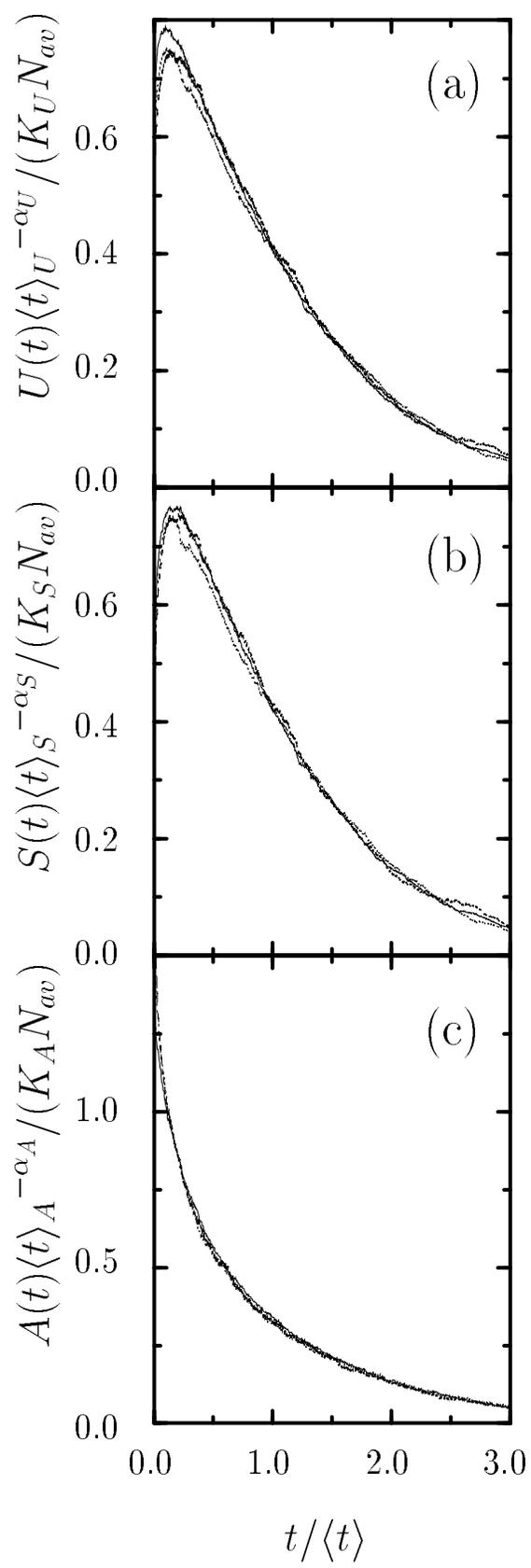
\begin{figure}
\caption{The scaling functions for the Manna four-state model for
$system size=128,256$ and $512$ (a) $f_U(\mu)$,  (b) $f_S(\mu)$ and (c) $f_A(\mu)$.
We observe that  the scaling functions $f_U(\mu)$ and $f_S(\mu)$
coincide while  $f_A(\mu)$ has a completely different form.}
\label{mannascale}
\end{figure}

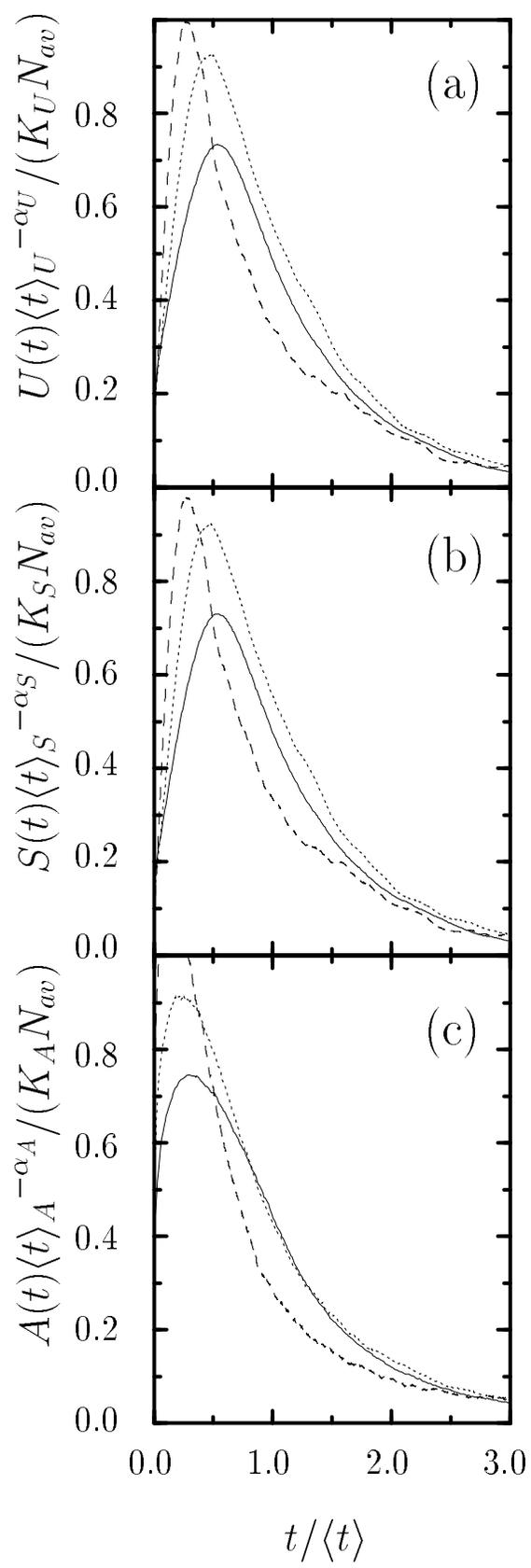
\begin{figure}
\caption{The functions (a) $f_U(\mu)$,  (b) $f_S(\mu)$ and (c) $f_A(\mu)$ 
for the Zhang model for
$system size=128,256$ and $512$.
We observe that functions obtained from different system sizes do not coincide, 
indicating that these are not scaling functions. This indicates that the Zhang model lacks
some of the characteristic features of a critical state found in the BTW and Manna models.
Interestingly, for each system size $f_U(\mu)$ and $f_S(\mu)$ still coincide, while  $f_A(\mu)$ 
is different.
}
\label{zhangscale}
\end{figure}

To obtain a more quantitative characterization of the terrace structure we chose typical large avalanches 
for each of the three models and plotted the terrace number $n$ as a function of its area (Fig.~\ref{terraces}).
The results for the Manna model are well fitted by an exponential,
while a nearly linear decay is observed for the BTW and Zhang models.  
 We have also measured, for the three models,
 the distribution exponents $\tau_a(n)$ for terraces no. $2,3,4$ and $5$ (Table~\ref{terracee}).
The results show quantitative differences in the avalanche structures, 
between the three models.
In all cases $\tau_a(n)$ decreases as $n$ is increased. The differences are significant with
the lowest $\tau_a(n)$ for Zhang, intermediate for BTW and highest for Manna models.
To obtain a more complete characterization of critical behavior in sandpile models we also
examine the time evolution of the energy, avalanche size and area during the avalanche, 
averaged over a large number of avalanches.
Combining results for system sizes $L=128,256$ and $512$ we draw the scaling functions
$f_U(\tau),f_S(\tau)$ and $f_A(\tau)$ that describe the averaged time evolution of the energy,
number of active sites, and area growth during the avalanche.
The scaling functions for the BTW, Manna and Zhang models are shown in 
Figs.~\ref{btwscale},~\ref{mannascale} and ~\ref{zhangscale}, respectively. 
For the BTW and Manna models we find excellent data collapse indicating 
scaling behavior. No such scaling is found for the Zhang model, indicating that it lacks
some of the features of a critical system, which are found in the BTW and Manna models.
For the BTW model (Figs~\ref{btwscale}) we observe that all the three scaling functions 
$f_U(\tau),f_S(\tau)$ and $f_A(\tau)$ are identical, so the system is basically described by a 
single scaling function.
For the Manna model, we find that $f_U(\tau)$ and $f_S(\tau)$ coincide, 
while   $f_A(\tau)$ is different.
For the Zhang model we find that there is no data collapse and there are no scaling scaling
function. Interestingly, for each system size the functions $f_U(\tau)$ and $f_S(\tau)$
are identical, while $f_A(\tau)$ is different.
The average times are found to depend on the system size according 
to $\langle t \rangle_X \sim L^{\beta_X}$ where  $X \in \{U, S, A\}$.
For the BTW model $\beta_U=1.51$,$\beta_S=1.43$ and $\beta_A=1.31$; for the
Manna model  $\beta_U=1.52$,$\beta_S=1.53$ and $\beta_A=1.48$; for the Zhang model one can
approximate these exponents by values  $\beta_U=1.5$,$\beta_S=1.46$ and $\beta_A=1.36$
but there is a significant deviation from a straight line.

\section{Summary and Conclusions}
	\label{summary}

We have studied universality in isotropic, abelian and non-abelian, sandpile models using a 
combination of extensive numerical simulations and an extended set of measures 
to characterize 
these models. In particular, we focused on the BTW  model (which is abelian, deterministic and isotropic),
Manna model (non-abelian, stochastic and isotropic on average) and the Zhang model 
(non-abelian, deterministic and isotropic). For each model we have calculated the critical exponents which
characterize an avalanche as a whole. These include the distribution exponents $\tau_x$, that
characterize the distribution of quantities such as avalanche size, area and lifetime, and the 
geometric exponents $\gamma_{xy}$, that relate the scaling properties of different quantities.
The geometric exponents $\gamma_{xy}$ are particularly useful for classification due to their
weak dependence on the system size. Comparing these exponents we find clear
indications that the BTW and Manna models belong to different universality classes, in 
agreement with previous simulations~\cite{B96}.
As for the Zhang model, the geometric critical exponents are not well defined. For all 
system size examined the functions $E[x|y]$ vs. $x$, where  $x \in \{s, a, t\}$, from
which the exponents  $\gamma_{xy}$ are obtained exhibit domains with different slopes
for small and large avalanches. The small avalanche behavior is similar to the 
BTW results, while the large avalanche behavior is different from both the BTW and 
Manna models.

The avalanche structures of the three models are found to be significantly different.
The BTW avalanche structure is the most regular, the Manna structure is the most
irregular and the Zhang avalanche structure is intermediate.

We have also examined measures of the dynamics during the avalanche. We found
 scaling functions for the time evolutions of the energy $f_U(\tau)$, number of active sites 
 $f_S(\tau)$, and the rate of area growth $f_A(\tau)$ in the BTW and Manna models. 
For the BTW model, all three scaling functions coincide, while for the Manna model
only the first two coincide. This is a qualitative different that further strengthen our 
conclusion that the two models belong to different universality classes.
For the Zhang model these functions do not exhibit scaling behavior.
This is a further indication that the Zhang model lacks some essential features of critical 
behavior, which appear in the BTW and Manna models, and thus belong to a 
different universality class. In fact, the only unambiguous scaling features of the Zhang
model are given by the distribution exponents $\tau_x$.
We thus conclude that the BTW, Manna and Zhang models belong to three universality 
classes.

Our results disagree with the conclusions of a number of recent studies. L\"{u}beck
studied the scaling behavior in the BTW and Zhang models using extensive numerical 
simulations~\cite{L97}.
Relying only on the distribution exponents he concluded that the BTW and Zhang models
belong to the same universality class. As we demonstrated above, the 
distribution exponents provide very limited characterization of the scaling behavior.
Therefore, these exponents alone do not provide enough information to support a
conclusion that two models belong to the same universality class.
Moreover, the distribution exponents are strongly dependent on the system size. One
should be careful in interpreting the results of the finite size analysis done in~\cite{L97},
based on an assumed size dependence which is not substantiated theoretically.
Corral and Diaz-Guilera derived nonlinear partial differential equations
based on the microscopic evolution rules of the BTW and Zhang models~\cite{C97}.
Using a dynamic RG approach they studied and concluded that the two models belong
to the same universality class. Vespignani, Zapperi and Pietronero used a real space RG 
approach and concluded that the BTW and Manna models belong to the same universality
class~\cite{V94}. The failure of these approaches to distinguish between the universality 
classes indicates that some key ingredient of the dynamics is not taken into account. 
We speculate the these ingredient may be related to symmetries such as the abelian
symmetry, as well as properties such as the deterministic vs. stochastic nature of the 
avalanche dynamics.

{\bf Acknowledgments}

We thank M. Paczuski for helpful discussions and correspondence.

\newpage

\begin{table}
\caption{The distribution exponent $\tau_a(n)$ for the areas of the terraces no. $n=2,3,4,5$, in a 2D
sandpile of size $128^2$.
For all three models, $\tau_a(n)$ tends to decrease as the terrace order $n$ increases.
The differences between the models are significant, with the lowest exponents for Zhang, intermediate
for BTW and highest for the Manna model.}
\label{terracee}
\end{table}

\newpage


\begin{table}
\centering
\begin{tabular}{cddd}
Exponent  &\multicolumn{3}{c}{model} \\ \cline{2-4}  
             & \multicolumn{1}{c}{BTW}   & \multicolumn{1}{c}
{Manna four-state}  
&\multicolumn{1}{c}{Zhang} \\ \hline

$\tau_a(2)$ &  2.05 $\pm$ 0.03 &  2.16  $\pm$ 0.03 & 1.98  $\pm$ 0.03    \\ 
$\tau_a(3)$ &  2.00 $\pm$ 0.04 &  2.11  $\pm$ 0.04 & 1.85  $\pm$ 0.04    \\ 
$\tau_a(4)$ &  1.95 $\pm$ 0.05 &  2.09  $\pm$ 0.05 & 1.77  $\pm$ 0.05    \\
$\tau_a(5)$ &  1.91 $\pm$ 0.05 &  2.00  $\pm$ 0.05 & 1.74  $\pm$ 0.05         
\end{tabular}
\end{table}

\newpage

\newpage

\end{document}